\documentclass[10pt,journal,letterpaper,twosides,twocolumns]{IEEEtran}

\usepackage{amsmath,amssymb,amscd,latexsym,dsfont,wasysym,bbm}
\usepackage{float}
\usepackage{color}
\usepackage{graphicx,subfigure,balance}
\usepackage{multirow,multicol}
\usepackage{verbatim}
\usepackage{psfrag}
\usepackage{comment}
\usepackage{makeidx}
\usepackage[]{algorithm}
\usepackage{algorithmic}
\usepackage{cite}

\usepackage{amsthm}

\newcommand{\tr}[1]{\textrm{#1}}
\newcommand{\mr}[1]{\mathrm{#1}}
\newcommand{\tnr}[1]{{\textnormal{#1}}}

\newcommand{\mc}[1]{\mathcal{#1}}
\newcommand{\mf}[1]{\mathsf{#1}}

\newcommand{\ms}[1]{\mathds{#1}}

\newcommand{\ov}[1]{\overline{#1}}

\newcommand{\ba}{\boldsymbol{a}}

\newcommand{\bd}{\boldsymbol{d}}

\newcommand{\bx}{\boldsymbol{x}}

\newcommand{\by}{\boldsymbol{y}}

\newcommand{\bz}{\boldsymbol{z}}

\newcommand{\figref}[1]{Fig.~\ref{#1}}

\newcommand{\secref}[1]{Sec.~\ref{#1}}

\newcommand{\appref}[1]{Appendix~\ref{#1}}

\newcommand{\exref}[1]{Example~\ref{#1}}
\newcommand{\propref}[1]{Proposition~\ref{#1}}

\newcommand{\lemref}[1]{Lemma~\ref{#1}}

\newcommand{\ie}{i.e.,~} 		
\newcommand{\eg}{e.g.,~}	

\newcommand{\argmax}{\mathop{\mr{argmax}}}
\newcommand{\argmin}{\mathop{\mr{argmin}}}

\newcommand{\set}[1]{\{#1\}}
\newcommand{\SET}[1]{\left\{#1\right\}}
\newcommand{\bset}[1]{\bigl\{#1\bigr\}}
\newcommand{\cd}{\cdot}
\newcommand{\ld}{\ldots}

\newcommand{\PR}[1]{\Pr\SET{#1}}       	
\newcommand{\pdf}{p}            			

\newcommand{\IND}[1]{\ms{I}[{#1}]}   	
\newcommand{\Ex}{\ms{E}}     			
\newcommand{\T}{^{\mr{T}}}            		
\newcommand{\dd}{\,\mr{d}}             		

\newcommand{\mcL}{\mc{L}}

\newcommand{\mcX}{\mc{X}}

\newcommand{\mfm}{\mf{m}}

\newcommand{\Real}{\mathbb{R}}		

\newcommand{\SNR}{\mathsf{snr}}  
\newcommand{\SNRrv}{\mathsf{SNR}}  
\newcommand{\SNRav}{\ov{\mathsf{snr}}}  
\newcommand{\X}{\mcX}	

\newcommand{\R}{R}              	
\newcommand{\Nb}{{\mathop{N_\tnr{b}}}} 	
\newcommand{\Ns}{{\mathop{N_\tnr{s}}}} 	

\newcommand{\Err}{\mf{ERR}}
\newcommand{\nack}{\mf{NACK}}  
\newcommand{\kmax}{K}		
\newcommand{\kk}{k}

\newcommand{\PER}{\mr{PER}}  
\newcommand{\DEC}{\mr{DEC}}  
\newcommand{\ENC}{\mr{ENC}}  

\usepackage[acronym,nonumberlist]{glossaries} 
\usepackage{hyperref}
\newacronym[\glsshortpluralkey=PDFs,\glslongpluralkey=probability density functions]{pdf}{PDF}{probability density function}
\newacronym[\glsshortpluralkey=CDFs,\glslongpluralkey=cumulative density functions]{cdf}{CDF}{cumulative density function}
\newacronym[\glsshortpluralkey=CCDFs,\glslongpluralkey=complementary cumulative density functions]{ccdf}{CDF}{complementary cumulative density function}
\newacronym[\glsshortpluralkey=PMFs,\glslongpluralkey=probability mass functions]{pmf}{PMF}{probability mass function}
\newacronym[]{lhs}{l.h.s.}{left-hand side}
\newacronym[]{rhs}{r.h.s.}{right-hand side} 

\newacronym[]{bicm}{BICM}{bit-interleaved coded modulation}
\newacronym[]{bicmid}{BICM-ID}{BICM with iterative demapping}
\newacronym[]{cm}{CM}{coded modulation}
\newacronym[]{tcm}{TCM}{trellis-coded modulation}
\newacronym[]{mlc}{MLC}{multi-level coding}
\newacronym[]{pam}{PAM}{pulse amplitude modulation}
\newacronym[]{bpsk}{BPSK}{binary phase shift keying}
\newacronym[]{qam}{QAM}{quadrature amplitude modulation}
\newacronym[]{psk}{PSK}{phase shift keying}
\newacronym[\glsshortpluralkey=LLRs,\glslongpluralkey=logarithmic likelihood ratios]{llr}{LLR}{logarithmic likelihood ratio}
\newacronym[]{map}{MAP}{maximum a posteriori}
\newacronym[]{ml}{ML}{maximum likelihood}
\newacronym[\glsshortpluralkey=MIs,\glslongpluralkey=mutual informations]{mi}{MI}{mutual information}
\newacronym[\glsshortpluralkey=GMIs,\glslongpluralkey=generalized mutual informations]{gmi}{GMI}{generalized mutual information}
\newacronym[]{eesm}{EESM}{exponential effective-SNR-mapping}
\newacronym[]{bicm-gmi}{BICM-GMI}{BICM generalized mutual information}
\newacronym[]{awgn}{AWGN}{additive white Gaussian noise}
\newacronym[]{amc}{AMC}{adaptive modulation and coding}
\newacronym[]{csi}{CSI}{channel state information}
\newacronym[]{cqi}{CQI}{channel quality indicator}
\newacronym[]{sp}{SP}{set-partitioning}
\newacronym[]{gsm}{GSM}{global system for mobile communications}
\newacronym[]{edge}{EDGE}{enhanced data rates for GSM evolution}
\newacronym[]{3gpp}{3GPP}{3rd generation partnership project}
\newacronym[]{dvb}{DVB}{digital video broadcasting}

\newacronym[\glsshortpluralkey=CCs,\glslongpluralkey=convolutional codes]{cc}{CC}{convolutional code}
\newacronym[\glsshortpluralkey=PCCCs,\glslongpluralkey=parallel concatenated convolutional codes]{pccc}{PCCC}{parallel concatenated convolutional code}
\newacronym[\glsshortpluralkey=TCs,\glslongpluralkey=turbo codes]{tc}{TC}{turbo code}
\newacronym{ldpc}{LDPC}{low-density parity-check}
\newacronym[]{ofdm}{OFDM}{orthogonal frequency-division multiplexing}
\newacronym[]{bep}{BEP}{bit-error probability}
\newacronym[]{wep}{WEP}{word-error probability}
\newacronym[]{sep}{SEP}{symbol-error probability}
\newacronym[]{pep}{PEP}{pairwise-error probability}
\newacronym[]{ttcm}{TTCM}{turbo-trellis coded modulation}
\newacronym[]{uep}{UEP}{unequal error protection}
\newacronym[\glsshortpluralkey=CENCs,\glslongpluralkey=convolutional encoders]{cenc}{CENC}{convolutional encoder}
\newacronym[]{mimo}{MIMO}{multiple-input multiple-output}
\newacronym[\glsshortpluralkey=SNRs,\glslongpluralkey=signal-to-noise ratios]{snr}{SNR}{signal-to-noise ratio}
\newacronym[]{msb}{MSB}{most-significative bit}
\newacronym[]{bcjr}{BCJR}{Bahl--Cocke--Jelinek--Raviv}
\newacronym[\glsshortpluralkey=SEDs,\glslongpluralkey=squared Euclidean distances]{sed}{SED}{squared Euclidean distance}
\newacronym[\glsshortpluralkey=EDs,\glslongpluralkey=Euclidean distances]{ed}{ED}{Euclidean distance}
\newacronym[\glsshortpluralkey=MEDs,\glslongpluralkey=minimum Euclidean distances]{med}{MED}{minimum Euclidean distance}
\newacronym[]{core}{CoRe}{constellation rearrangement}
\newacronym[]{msd}{MSD}{multistage decoding}
\newacronym[]{pdl}{PDL}{parallel decoding of the individual levels}
\newacronym[\glsshortpluralkey=GCs,\glslongpluralkey=Gray codes]{gc}{GC}{Gray code}
\newacronym[]{brgc}{BRGC}{binary-reflected Gray code}
\newacronym[]{nbc}{NBC}{natural binary code}
\newacronym[]{fbc}{FBC}{folded-binary code}
\newacronym[]{bsgc}{BSGC}{binary semi-Gray code}
\newacronym[]{msp}{MSP}{modified set-partitioning}
\newacronym[]{ssp}{SSP}{semi set-partitioning}
\newacronym[]{fhd}{FHD}{free Hamming distance}
\newacronym[]{mfhd}{MFHD}{maximum free Hamming distance}
\newacronym[]{ods}{ODS}{optimal distance spectrum}
\newacronym[]{iud}{i.u.d.}{independent and uniformly distributed}
\newacronym[]{ud}{u.d.}{uniformly distributed}
\newacronym[]{iid}{i.i.d.}{independent, identically distributed}
\newacronym[]{bico}{BICO}{binary-input continuous-output}
\newacronym[]{gh}{GH}{Gauss--Hermite}

\newacronym[\glsshortpluralkey=BSs,\glslongpluralkey=base-stations]{bs}{BS}{base-station}
\newacronym[\glsshortpluralkey=MSs,\glslongpluralkey=mobile-stations]{ms}{MS}{mobile-stations}

\newacronym[]{phy}{PHY}{physical layer} 
\newacronym[]{llc}{LLC}{logical link control} 
\newacronym[]{fft}{FFT}{fast Fourier transform} 
\newacronym[]{cf}{CF}{characteristic function} 
\newacronym[]{mgf}{MGF}{moment generating function} 
\newacronym[]{ee}{EE}{energy efficiency} 
\newacronym[]{kkt}{KKT}{Karush--Kuhn--Tucker} 
\newacronym[]{mcs}{MCS}{modulation/coding scheme} 
\newacronym[]{fec}{FEC}{forward error correction}
\newacronym[]{arq}{ARQ}{automatic repeat request}
\newacronym[]{harq}{HARQ}{hybrid ARQ}
\newacronym[]{tarq}{TARQ}{truncated HARQ}
\newacronym[]{rrharq}{RR-HARQ}{repetition redundancy HARQ}
\newacronym[]{irharq}{IR-HARQ}{incremental redundancy HARQ}
\newacronym[]{ack}{ACK}{positive acknowledgment}
\newacronym[]{nack}{NACK}{negative acknowledgment}
\newacronym[]{crc}{CRC}{cyclic redundancy check}
\newacronym[]{dp}{DP}{dynamic programming}
\newacronym[]{gp}{GP}{geometric programming}
\newacronym[]{per}{PER}{packet error rate}
\newacronym[]{op}{OP}{outage probability}
\newacronym[]{spa}{SPA}{saddle-point approximation}
\newacronym[]{mrc}{MRC}{maximum ratio combining}
\newacronym[]{mdp}{MDP}{Markov decision process}

\newtheorem{proposition}{Proposition}

\newtheorem{lemma}{Lemma}
\newtheorem{example}{Example}

\begin{document}

\title{Modelling Decoding Errors in HARQ}
\author{Redouane Sassioui, Etienne Pierre-Doray, Leszek Szczecinski, and Benoit Pelletier%
\thanks{%
R. Sassioui and L. Szczecinski are with INRS-EMT, Montreal, Canada. [e-mail: \{redouane.sassioui,~leszek\}@emt.inrs.ca].}%
\thanks{
E. Pierre-Doray is with  INRS-EMT and Polytechnique de Montreal, Canada 
[e-mail: etipdoray@gmail.com].}
\thanks{%
Benoit Pelletier is with InterDigital Communications, Montreal, Canada. [e-mail: Benoit.Pelletier@interdigital.com].} 
\thanks{%
The work was supported by NSERC, Canada under \emph{Undergraduate Student Research Awards} program and by FRQNT, Quebec, under \emph{Recherche en  \'equipe} grant.}%

}

\maketitle
\thispagestyle{empty}

\begin{abstract}
In this work we address the issues of probabilistic modelling of the decoding errors in \gls{harq} rounds. In particular we i)~claim that the assumption of independence of decoding errors, used implicitly in various works on this subject, is an approximation, and ii)~propose equally simple but much more accurate method to calculate the probability of the sequence of decoding errors.
The model we propose is useful from the point of view of performance evaluation, system-level simulation, and/or link adaptation. Its simplicity leads also to  closed form expression for the outage probability and for the average number of transmissions in block-fading channel. 
\end{abstract}

\section{Introduction}\label{Sec:Intro}
Modern wireless systems use \gls{harq} protocols to deal with unavoidable transmission errors in noisy and unpredictably varying channels. \gls{harq} is a ``handshaking'' protocol where the receiver uses a feedback channel to inform the transmitter about a successful decoding of the transmitted message via a \gls{ack}. 

To deliver a message, \gls{harq} relies both on channel coding and on retransmissions which are carried out in response to the decoding errors. The latter are indicated by a \gls{nack}, which triggers a new \emph{\gls{harq} round}. During each round a coded message is transmitted. Multiple \gls{harq} rounds may be necessary to deliver the message and they stop upon a reception of \gls{ack}, when the transmitter sends a new encoded message.

In order to characterize the performance of the \gls{harq}, the probabilistic model of decoding errors has to be known; in this work we propose such a model that is simple and accurate, and we compare it agains the existing alternatives.

In this work, we focus on \gls{rrharq} (sometimes referred to as \emph{Chase combining \gls{harq}} \cite{Chase85}), where  the same codeword is transmitted (\ie repeated) in each round. 
Receiver in \gls{rrharq} performs a \gls{mrc} of the signals received in various rounds, and the probability of the decoding error in the $\kk$th round, $\PR{\Err_{\kk}}$, depends then on the accumulated \gls{snr} (the sum of \gls{snr}s observed in $\kk$ subsequent rounds). However,  when analyzing \gls{harq}, we are mostly interested in the probability of the errors sequence $f_{\kk}\triangleq\PR{\Err_1, \ld, \Err_{\kk}}$ which, in general, depends not only on the accumulated \gls{snr}s but also on the \gls{snr}s in all $\kk$ rounds.

We need to calculate the errors sequence probability $f_{\kk}$ in various cases such as 
\begin{itemize}
\item \textbf{Performance evaluation}, where we must evaluate  $f_{\kk}$ in order to find the important parameters of the \gls{harq}-based transmission, such as the throughput or the average delay \cite{Caire01}.

\item \textbf{System-level simulation}, where, in order to model the behaviour of systems with many users,  a prohibitively complex decoding of users' packets is replaced by a probabilistic model of the decoding errors, which are then generated pseudo-randomly according to the probability $\PR{\Err_\kk|\Err_{\kk-1},\ld, \Err_1}=f_\kk/f_{\kk-1}$ \cite[Sec.~A.2.2]{3gpp25.896_V6.0.0}\cite{Latif13}.

\item \textbf{Link adaptation} where resources (such as power or bandwidth) are optimized to attain the target performance often defined in terms of the conditional probability $f_\kk/f_{\kk-1}$ or the final communication failure probability $f_\kk$. In order to fulfill the requirement, and adequately assign resources (\eg the power) the transmitter needs to predict the probability of decoding error.
\end{itemize}

The straightforward solution of calculating the errors sequence probability $f_{\kk}$ consists in storing each conditional probability $\PR{\Err_l|\Err_{l-1},\ld,\Err_1}$ in a multidimensional table indexed with \gls{snr}s, $\SNR_1,\ld,\SNR_l$. Such an approach was proposed for system level simulations in \cite[Sec.~A.2.2]{3gpp25.896_V6.0.0}, however, it provides little insight into the functionality of the \gls{harq} protocols where the analytical prediction of the error probability is much more useful.

To this end, a simple analytical model which assumes the threshold decoding is often used: it assumes that the error is declared only if the accumulated \gls{snr} is below the decoding threshold \cite{Caire01,Szczecinski13,Larsson14}. In this case, knowing the decoding threshold and the \gls{snr}s suffices to calculate the probability $f_{\kk}$ \cite{Caire01,Larsson14}. 

However, while the threshold-decoding model is important from an information-theoretical point of view, it fails to capture the characteristics of practical encoders/decoders operating with finite-length codewords. This problem already gained attention and several previous works analyzed \gls{harq} protocols operating with such ``imperfect'' decoders, \eg \cite{Liu04,KimHan04,Zheng05,Cho06,Femenias09,Lagrange10,Ramis11,Lagrange11,Mukhtar13,Ge14,Park15}. The common idea of these  works was to exploit the function relating the probability of decoding error to the \gls{snr} in a one-shot transmission (\ie without retransmissions). Such a \gls{per} function still has to be found by simulations/measurements but since this function is scalar the problem is  greatly simplified.

In this work, we investigate how the \gls{per} function should be used to calculate the errors sequence probability. Despite its simplicity and fundamental importance for the applications we mentioned, this question has not been yet addressed up front and the existing implicit models are  inaccurate. 

The main contributions of this work is to propose a new model of \gls{harq} decoding errors (the so-called deterministic errors model), which is an accurate and formal upper bound on the error sequence probability $f_{\kk}$. We also show that the model suggested in \cite{ZhengViswanthan03} and then used in\cite{KimHan04,Zheng05,Cho06,Lagrange10,Lagrange11,Mukhtar13,Ge14,Park15} is based on the implicit assumption of errors independence and may severely underestimate the value of $f_{\kk}$. Furthermore, we i)~evaluate the bounding error in our model and ii)~show how to use it to evaluate the average outage probability of receivers operating without \gls{csi} in \gls{iid} block-fading channel.

The remainder of the paper is organized as follows: in \secref{Sec:Model} we define the operation of \gls{rrharq}, explain the objective of the work, and present the error model used in the literature. A new model of the decoding errors is then proposed in \secref{Sec:dec.failure}. Analysis and examples of applications are shown in \secref{Sec:Numerical}, and we conclude the work in \secref{Sec:Conclusions}.

\section{System model}\label{Sec:Model}

Consider the transmission system where the information message $\mfm\in\set{0,1}^{\Nb}$ is encoded into a codeword $\bx=\ENC[\mfm]\in\mcX^\Ns$, where $\mcX$ is a complex constellation. The \emph{nominal} transmission rate is given by $\R=\Nb/\Ns$.

The communication may be carried in multiple \emph{rounds}. The transmission in each round produces the outcome
\begin{align}\label{}
\by_\kk=\sqrt{\SNR_\kk}\bx+\bz_\kk, 
\end{align}
where $\bz_\kk$ models a Gaussian noise with zero-mean and unit variance and, modelling symbols in $\bx$ as zero-mean unit-variance random variables, $\SNR_\kk$ is the \gls{snr} experienced at the receiver at the $k$th round.

Since the same codeword $\bx$ is transmitted in each round, the decoding result may be obtained from the all channel outcomes processed via \gls{mrc}
\begin{align}
\label{mrc.by}
\by_{[\kk]}&\triangleq\frac{1}{\sqrt{\SNR_{[\kk]}}}\sum_{l=1}^k\sqrt{\SNR_l}\by_l\\
\label{mrc.by.2}
&=\sqrt{\SNR_{[\kk]}} \bx+\bz_{[\kk]},
\end{align}
where
\begin{align}\label{bz.kk}
\bz_{[\kk]}=\frac{1}{\sqrt{\SNR_{[\kk]}}}\sum_{l=1}^k\sqrt{\SNR_l}\bz_l
\end{align}
 is a unit variance Gaussian ``effective'' noise,  and
\begin{align}\label{SNR.sum.k}
\SNR_{[\kk]}=\sum_{l=1}^k\SNR_l,
\end{align}
is the accumulated \gls{snr}; the \gls{mrc} operation  in \eqref{mrc.by} is also referred to as \emph{Chase combining} \cite{Chase85,Caire01}.


\subsection{Error events}
The result of the decoding after the $\kk$th round is based on channel outcomes in  rounds $1,\ld,\kk$, $\hat{\mfm}_\kk=\DEC_\kk[\by_1,\ld,\by_\kk]$, and the decoding error is defined as 
\begin{align}\label{Err.k}
\Err_\kk\triangleq\set{\hat{\mfm}_\kk\neq \mfm}.
\end{align}

If $\Err_k$ occurs, the receiver sends a \gls{nack}, which triggers the next round. This continues untill the maximum allowed number of rounds, $\kmax$, is reached or untill there is no error in the $\kk$th round, \ie $\hat{\mfm}_\kk=\mfm$, which is confirmed with an \gls{ack} sent by the receiver.

Due to \eqref{mrc.by.2} the decoding based on $\by_1,\ld,\by_\kk$ reduces to 
\begin{align}\label{dec.k}
\hat{\mfm}_k=\DEC[\by_{[\kk]}],
\end{align}
\ie the probability of decoding error may be characterized by the same \gls{per} function in each transmission
\begin{align}\label{PER.def}
\PR{\Err_\kk }=\PER(\SNR_{[\kk]}).
\end{align}

That is, even if the function $\PER(\cd)$ depends on the encoding and decoding algorithms (and must be, in general, obtained by simulations/measurements), it is affected by the past channel \gls{snr}s $\SNR_1,\ld, \SNR_\kk$ only via accumulated \gls{snr}, $\SNR_{[\kk]}$.

We also define the communication \emph{failure} after $\kk$ rounds through the errors-sequence
\begin{align}\label{nack.k}
\nack_\kk \triangleq \set{\Err_\kk, \Err_{\kk-1}, \ld, \Err_1},
\end{align}
which is the event after which the $k$th \gls{nack}  is emitted by the receiver.

The probability of the communication failure is then given by
\begin{align}
\label{f.k}
f_\kk&=\PR{\nack_\kk}\\
\nonumber
&=\PR{\Err_\kk,\Err_{\kk-1},\ld,\Err_1}\\
\label{f.k.cond}
&=\PR{\Err_{\kk}|\nack_{\kk-1}}f_{\kk-1}.
\end{align}

This formulation can be used, for example, in system-level simulations where the error in the transmission of the $k$th round is generated using the  probability
\begin{align}\label{E.cond}
\PR{\Err_\kk | \nack_{\kk-1}}=\frac{f_{\kk}}{f_{\kk-1}}.
\end{align}

To calculate $f_\kk$ we need to evaluate the joint distribution of the errors events, $\Err_1,\ld, \Err_{\kk}$ which, in general, is not known and difficult to acquire. Instead, we want to approximate \eqref{f.k} only using  the known \gls{per} function $\PER(\cd)$ defined in \eqref{PER.def}.

\subsection{Independent error model}

The following expression for the probability of communication failure after the transmission of the $\kk$th round was used in \cite[Eq.~(5)]{ZhengViswanthan03}\footnote{It was used there as part of the expression for the  throughput.}
\begin{align}\label{f.k.IE}
\tilde{f}_\kk&\triangleq \prod_{l=1}^k \PR{\Err_l}\\
&=\prod_{l=1}^k \PER(\SNR_{[l]}).
\end{align}

We note that \eqref{f.k.IE} may be derived from \eqref{f.k.cond}, using \eqref{PER.def}, if we assume that the decoding errors are independent, \ie
\begin{align}\label{Pr.k.approx}
\PR{\Err_\kk |\nack_{\kk-1}}= \PR{\Err_\kk}.
\end{align}

Under this ``independent error'' (IE) model, \eqref{E.cond} becomes
\begin{align}\label{E.cond.IE}
\PR{\Err_\kk | \nack_{\kk-1}}\approx\frac{\tilde{f}_{k}}{\tilde{f}_{\kk-1}}= \PER(\SNR_{[\kk]}).
\end{align}

\section{Deterministic error model}\label{Sec:dec.failure}

Since IE model was adopted in \cite{ZhengViswanthan03} without discussion and was further reused in other works, such as\cite{KimHan04,Zheng05,Cho06,Lagrange10,Lagrange11,Mukhtar13,Ge14,Park15}, we   emphasize that \eqref{f.k.IE} is strictly an approximation, that is $f_\kk\approx \tilde{f}_\kk$, which we formalize in the following.

\begin{proposition}
The decodings errors $\set{\Err_l}_{l=1}^\kk$ are not independent.
\begin{proof}
For the proof it is enough to demonstrate existence of the operating conditions $\SNR_1,\ld,\SNR_\kk$ under which two errors \eg $\Err_\kk$ and $\Err_{\kk-1}$, where $\kk>1$, are not independent, that is, for which the condition $\PR{\Err_{\kk-1},\Err_{\kk}}=\PR{\Err_{\kk-1}}\PR{\Err_{\kk}}$ \emph{does not} hold.

Let us suppose that $\SNR_{\kk}=0$. Then $\by_{[\kk]}=\by_{[\kk-1]}$, see \eqref{mrc.by}, and thus $\Err_{\kk-1}\implies\Err_{\kk}$ and then
\begin{align}\nonumber
\PR{\Err_{\kk-1},\Err_{\kk}}&=\PR{\Err_{\kk-1}}\\
\nonumber
& > \PR{\Err_{\kk}}\PR{\Err_{\kk-1}}\\
&=\big(\PR{\Err_{\kk-1}}\big)^2.
\end{align}
%
%
%
\end{proof}
\end{proposition}

In more general terms, the dependance between the events $\Err_{\kk-1}$ and $\Err_{\kk}$ is caused by the correlation of the effective noises $\bz_{[\kk-1]}$ and $\bz_{[\kk]}$ which is apparent when transforming \eqref{bz.kk} as follows
\begin{align}\label{z.kk.recursive}
\sqrt{\SNR_{[\kk]}}\bz_{[\kk]}=\sqrt{\SNR_{\kk}}\bz_\kk+ \sqrt{\SNR_{[\kk-1]}}\bz_{[\kk-1]}.
\end{align}

In other words, knowing that error occurred in the round $\kk-1$, we have a prior knowledge about $\bz_{[\kk-1]}$ which affects the distribution of $\bz_{[\kk]}$ via \eqref{z.kk.recursive}. This happens due to the fact that the decoder in the $\kk$th round uses the same channel outcomes $\by_1,\ld,\by_{\kk-1}$ as in the $(\kk-1)$th round. This phenomenon is emphasized by the choice of the values of \gls{snr}s we used in the example in the proof, however, in general, the dependence is not caused by the relationship between the \gls{snr}s or by a specific channel model.

{\bf Proposed model}

The model we propose now originates from the following heuristics: if the decoding based on $\by_{[\kk]}$ fails, the decoding based on $\by_{[\kk-1]}$ is likely to fail as well because the accumulated \gls{snr}  satisfies $\SNR_{[\kk]}>\SNR_{[\kk-1]}$, in other words we postulate the following relationship
\begin{align}\label{Err.joint.DE.2}
\PR{ \Err_1, \ld, \Err_{\kk-1}|\Err_\kk}&\approx 1\\
\label{Err.joint.DE.3}
\PR{\Err_1, \ld, \Err_{\kk-1},\Err_\kk }&\approx \PR{\Err_\kk};
\end{align}
thus, the errors $\Err_1,\ld,\Err_{\kk-1}$ may be treated as (semi) deterministic events once $\Err_\kk$ is observed.  Such a \emph{deterministic} error (DE) model is still an approximation because we do not know the joint distribution of the error events $\Err_1,\ld, \Err_k$.

Here, it may be tempting to remove the approximation sign from \eqref{Err.joint.DE.2} using the following reasoning: ``if the error occurs in the round $\kk$ it must have occurred also in the previous rounds (otherwise the $\kk$th round would have not taken place).'' 

While this reasoning indeed applies to the communication failure events (which are the series of decoding errors, see \eqref{nack.k},  and thus relate to the operation of the \gls{harq} protocol), it does not apply to the decoding errors which merely depend on the operation of the decoder. To understand it better, a convenient way of thinking is to consider that $\by_{l}, l=1,\ld,k$ are available \emph{simultaneously}. This  is useful because it allows us  to consider the relationship between the errors $\set{\Err_l}_{l=1}^\kk$ in abstraction of the indexing related to the rounds (which occur successively). 

To emphasize again the difference between the events $\nack_k$ (the errors sequence)  and $\Err_k$ (the decoding error) we consider the following example.

\begin{example}[``Backward'' conditional probability]\label{ex:backward.errors}
A simple consequence of the definition \eqref{nack.k} is that the joint probability of communication failures writes as $\PR{\nack_{\kk-1}, \nack_{\kk}}=\PR{\nack_{\kk}}$; the expression for the conditional probability is thus given by $\PR{\nack_{\kk-1}| \nack_{\kk}}=1$ and can be verbalized as `` to receive a \gls{nack}  in the $\kk$th round, we must have received a \gls{nack}  in the round $(\kk-1)$''.

We can interpret now the errors $\Err_\kk$ and $\Err_{\kk-1}$ as results of decoding  based on $\by_{[\kk]}$  and $\by_{[\kk-1]}$ (which are  available simultaneously). This  is indeed useful because it is possible to have error when decoding using $\by_{[\kk]}$ and decode correctly using $\by_{[\kk-1]}$;  thus,  $\PR{\Err_{\kk-1}|\Err_{\kk}}\neq1$. In fact, to conclude anything about the conditional probability $\PR{\Err_{\kk-1}|\Err_{\kk}}$ we need to know the joint probability $\PR{\Err_\kk,\Err_{\kk-1}}$.
\end{example}

From \eqref{Err.joint.DE.3} we obtain the approximation $f_\kk\approx\hat{f}_\kk$ with
\begin{align}
\label{f.k.DE}
\hat{f}_\kk &\triangleq\PR{\Err_\kk}\\
&=\PER(\SNR_{[\kk]}).
\end{align}

From the simple relationship 
\begin{align}\label{Pr.nack.bound}
\PR{\Err_1,\ld,\Err_\kk } &\leq \PR{\Err_k}\\
f_{\kk}&\leq \hat{f}_{\kk}.
\end{align}
we conclude that $\hat{f}_\kk$ is a formal upper bound on $f_\kk$.

Then, adopting the DE model, for the system-level simulations we have to use
\begin{align}\label{E.cond.DE}
\PR{\Err_\kk | \nack_{\kk-1}} \approx \frac{\hat{f}_{\kk}}{\hat{f}_{\kk-1}}= \frac{\PER(\SNR_{[\kk]})}{\PER(\SNR_{[\kk-1]})},
\end{align}
which is very different from \eqref{E.cond.IE} and their comparison in made in \secref{Sec:Numerical}.

\begin{example}\label{Ex:system.sim}
We assume that $\kk-1$ rounds were carried out, the accumulated \gls{snr} is given by $\SNR_{[\kk-1]}$, and the event $\nack_{\kk-1}$ is observed (communication failure). We further assume that in the $\kk$th round, we observe a very weak (null) \gls{snr}, \ie $\SNR_\kk= 0$ and thus, $\by_{[\kk]}=\by_{[\kk-1]}$, see \eqref{mrc.by}, and $\SNR_{[\kk]}=\SNR_{[\kk-1]}$, see \eqref{SNR.sum.k}.

Before evoking the approximations let us see what will be the decoding result after the $\kk$th round. This is, in fact, very simple: because $\SNR_\kk=0$, the $\kk$th round observation, $\by_{\kk}$, does not contribute any new information and thus, since decoding based on $\by_{[\kk-1]}$ failed, it must also fail when using the same signal $\by_{[\kk]}=\by_{[\kk-1]}$. Consequently, we obtain $\PR{\Err_\kk|\nack_{\kk-1}}=1$.

We can now compare IE and DE models to see how they predict the probability of decoding error in our example: using DE model \eqref{f.k.DE} we obtain $\PR{\Err_\kk | \nack_{\kk-1}}\approx \PER(\SNR_{[\kk-1]})/\PER(\SNR_{[\kk-1]})=1$, which is an exact value. On the other hand, using IE model \eqref{E.cond.IE} we obtain $\PR{\Err_\kk | \nack_{\kk-1}}\approx \PER(\SNR_{[\kk-1]})$ which, depending on the value of $\SNR_{[\kk-1]}$ may be very optimistic.
\end{example}

\begin{example}[Idealized threshold decoding]
In the threshold decoding assumption, we declare the error occurs if and only if the accumulated \gls{snr} exceeds the decoding threshold $\SNR_\tnr{th}$, thus
\begin{align}\label{PER.binary}
\PER(\SNR)=\IND{\SNR<\SNR_\tnr{th}},
\end{align}
where $\IND{\cd}$ is the indicator function. 

This idealized assumption is often used to analyze \gls{harq}, \eg \cite{Caire01,Szczecinski13,Larsson14}. Since we know that $\SNR_{[\kk]}\geq\SNR_{[\kk-1]}\geq \ld \geq \SNR_{[1]}$, then 
\begin{align}
\tilde{f}_\kk&=\prod_{l=1}^k \PER(\SNR_{[l]})\\
&= \PER(\SNR_{[\kk]})\\
&=\hat{f}_\kk
\end{align}
and thus, under the threshold decoding assumption, both IE and DE models produce identical results.
\end{example}

\section{Comparison of the models}\label{Sec:Numerical}

Our objective now is to assess the accuracy of the approximations $\tilde{f}_k$ and $\hat{f}_k$ provided, respectively by the IE and DE models, comparing them against the exact value of failure probability $f_k$. 

\subsection{Analytical insight: ML decoding}\label{Sec:analytical}

Calculation of the decoding error probability depends on the encoding and the decoding algorithm and, in general, resists the analytical efforts. This difficultly is alleviated if we assume that the decoder applies the \gls{ml} principle $\DEC[\by]=\argmax_{\mfm} \pdf (\by_{[\kk]}|\ENC[\mfm]) =\argmin_{\mfm} \| \by_{[\kk]}-\sqrt{\SNR_{[\kk]}}\ENC[\mfm]\|$.

We can then use the well-known union bound \cite[Sec.~6.2.1]{Szczecinski_Alvarado_book}
\begin{align}\label{Pr.err.bound}
\PR{\Err_\kk}&< \frac{1}{2^{\Nb}}
\sum_{\substack{\bx,\bx'\\\bx\neq\bx'}} 
\PR{\bx\xrightarrow{[\kk]}\bx'},
\end{align}
where \mbox{$\bset{\bx\xrightarrow{[\kk]}\bx'}$} denotes a pairwise-error event, \ie where $\bx'$ is more likely than $\bx$,\footnote{This event is conditioned on  the codeword $\bx$ being transmitted; we leave this condition implicit to alleviate the notation.} that is
\begin{align}
\nonumber
\bset{\bx\xrightarrow{[\kk]}\bx'}&=\bset{ \|\by_{[\kk]} -\sqrt{\SNR_{[\kk]}}\bx' \|< \| \by_{[\kk]}-\sqrt{\SNR_{[\kk]} }\bx  \|  }\\
\nonumber
&= \set{ \|\sqrt{\SNR_{[\kk]}}\bd+\bz_{[\kk]}\|<\| \bz_{[\kk]} \|}  \\
\label{snr.v.kk}
&= \SET{\tfrac{1}{2}\sqrt{\SNR_{[\kk]}}\|\bd\|<v_{[\kk]} },
\end{align}
where $\bd=\bx-\bx'$,
\begin{align}\label{v.kk}
v_{[\kk]}&=-\bd^\tr{H}\bz_{[\kk]}/\|\bd\|
\end{align}
is a zero-mean, unit-variance Gaussian variable, and $\bd^\tr{H}$ is the conjugate transpose of $\bd$. 

Thus we obtain 
\begin{align}\label{Pr.err.bound.2}
\PR{\Err_\kk}&
<\frac{1}{2^{\Nb}}\sum_{d=d_\tr{free}}^\infty C_d P_{\kk}(d)
\end{align}
where $C_{d}$ is the Euclidean distance spectrum of the code \cite[Sec.~6.2.3]{Szczecinski_Alvarado_book}, $d_\tr{free}$ is the minimum Euclidean distance between two distinct codewords, and 
\begin{align}\label{}
P_{\kk}(d)&=\PR{ 0.5\sqrt{\SNR_{[\kk]}} d<v_{[\kk]} }.
\end{align}
is the \gls{pep} for the codewords in the Euclidean distance$\|\bd\|=d$.

Similarly, we can approximate
\begin{align}\label{Pr.err.err.bound}
\PR{\Err_1\wedge\ld\wedge \Err_{\kk-1} \wedge \Err_{\kk}}< \sum_{d=d_\tr{free}}^\infty 
C_{d}   P_{1:\kk}(d),
\end{align}
where 
\begin{align}
\nonumber
P_{l:\kk}(d)\triangleq &\PR{ \bx\xrightarrow{[l]}\bx' \wedge \ld\wedge \bx\xrightarrow{[\kk-1]}\bx' \wedge \bx\xrightarrow{[\kk]}\bx' }\\
\nonumber
= 
&\Pr\Big\{ 
\tfrac{1}{2}\sqrt{\SNR_{[l]}}d<v_{[l]}
\wedge\ld\wedge\tfrac{1}{2}\sqrt{\SNR_{[\kk-1]}}d<v_{[\kk-1]}\\
\label{P.kk}
&\qquad~\quad
\wedge 
\tfrac{1}{2}\sqrt{\SNR_{[\kk]}}d<v_{[\kk]}
\Big\}.
\end{align}

Due to \eqref{Pr.err.bound.2} and \eqref{Pr.err.err.bound}, comparing $\PR{\Err_1,\ld, \Err_{\kk}}$ with its approximations $\PR{\Err_\kk}$ (resulting from the DE model) may be done by a comparison of the respective \gls{pep}s,  $P_{1:\kk}(d)$ and $P_{\kk}(d)$. The comparison will allow us to highlight the conditions under which these models are accurate. 

\begin{proposition}[PEP bounds]\label{Prop1} The \gls{pep} may be limited as follows
\begin{align}
\label{property.1}
&\frac{1}{2^{\kk-1}} P_\kk(d) \leq P_{1:\kk}(d)\leq P_\kk(d),
\end{align}
\begin{proof}
\appref{Appendix.proof}.
\end{proof}
\end{proposition}

\begin{proposition}\label{Corr1} 

The upper and lower bounds in \eqref{property.1} are attainable under the following conditions
\begin{itemize}
\item 
If there exist $l$ such that $\SNR_l>0$ and $\forall l'\neq l,  \SNR_{l'}=0$ then the upper bound is attained
\begin{align}
\label{property.2}
P_{1:\kk}(d)= P_{\kk}(d).
\end{align}
\item 
For $t_l\triangleq \SNR_{l}/\SNR_{l-1}, 2\leq l\leq\kk$, then
\begin{align}\label{}
\label{property.3}
&\lim_{t_2,\ld,t_\kk \rightarrow \infty} P_{1:\kk}(d)= \frac{1}{2^{\kk-1}}P_{\kk}(d).
\end{align}
\end{itemize}
\end{proposition}

The proof of \propref{Corr1} is tedious and does not provide further insight into the work; we thus omitted it for brevity.

The result \eqref{property.2} is quite simple and should be expected using reasoning shown in \exref{Ex:system.sim}. On the other hand, \eqref{property.3} provides a strong lower bound on the \gls{pep} $P_{1:\kk}(d)$ in terms of $P_\kk(d)$. While this bound becomes relatively loose for large $\kk$, it is very useful for $\kk=2$ and $\kk=3$ which are common values in the practical setup.

\begin{figure}[tb]
\newcommand{\scalefig}{0.9}
\psfrag{xlabel}[ct][ct][\scalefig]{$t_2$}
\psfrag{xlabelc}[ct][ct][\scalefig]{$t_3$}
\psfrag{ylabel}[ct][ct][\scalefig]{PEP}
\psfrag{PEP1k}[cl][cl][\scalefig]{$P_{1:\kk}(d)$}
\psfrag{PEPk}[cl][cl][\scalefig]{$P_\kk(d)$}
\psfrag{PEPk/2k}[cl][cl][\scalefig]{$\frac{1}{2^{\kk-1}}P_{\kk}(d)$}
\psfrag{PEPk/2}[cl][cl][\scalefig]{$\frac{1}{2}P_{\kk}(d)$}
\psfrag{k=2}[cl][cl][\scalefig]{$\kk=2$}
\psfrag{k=3}[cl][cl][\scalefig]{$\kk=3$}
\centering
\scalebox{1}{\includegraphics[width=\linewidth]{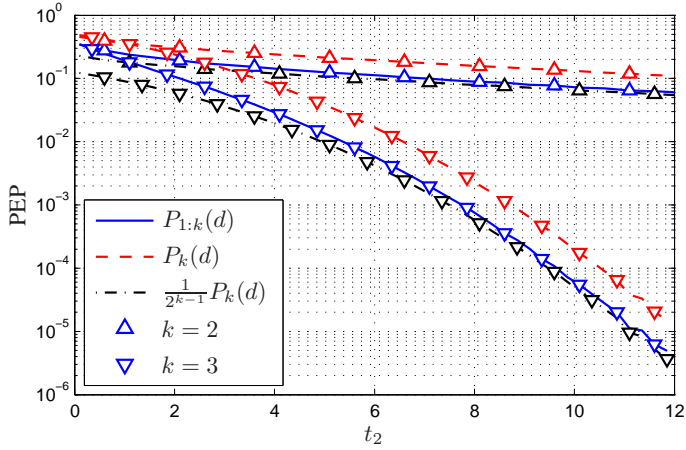}}\\
\caption{$P_{1:\kk}(d)$ and $P_\kk(d)$  as a function of $\set{t_l}_{l=2}^\kk$, where $t_l=t_2$, $d=1$ and  $\SNR_{1}=-3$dB. }\label{Fig:PEP}
\end{figure}

To illustrate their impact on $P_{1:\kk}(d)$ we show in \figref{Fig:PEP} the behaviour of  $P_{1:\kk}$ as a function of $t_l=t_2$.  We observe that the limit \eqref{property.3} is practically satisfied for $t_l>2$ if $\kk=2$; this means $3$dB difference between $\SNR_1$ and $\SNR_2$ is sufficient to attain the bound. On the other hand,  large values of $t_l$ are required to attain the lower bound for $\kk=3$.

We can also expect that using $P_{1:\kk}(d)\approx \prod_{l=1}^{\kk} P_l(d) $ (which may be seen as a proxy of the IE model) must provide very poor (optimistic) results, especially for small values of $P_l(d)$. This conjecture will be validated by the simulation examples.

\subsection{Simulations: Validation with practical codes}

Simulations will now be used to asses the accuracy of the analyzed approximations as well as to highlight the bounds obtained in \propref{Prop1}.

We use the message of $\Nb=512$ bits, and i)~the rate-$1/3$ turbo encoder (two parallel convolutional encoders with generator polynomials $[15/13]$) and the the max-log \gls{map} iterative decoder (with four iterations), and  ii)~rate-$1/2$ convolutional encoder with generator polynomials $[1, 15/13]$ and a Viterbi decoder. The code-bits are used to modulate \gls{bpsk} symbols, \ie $\X=\set{-1,1}$.

\begin{figure}[tb]
\newcommand{\scalefig}{0.9}
\psfrag{xlabelT}[ct][ct][\scalefig]{$\SNR_{[2]}$[dB]}
\psfrag{xlabelC}[ct][ct][\scalefig]{$\SNR_{[3]}$[dB]}
\psfrag{f2}[cl][cl][\scalefig]{$\hat{f}_{2}$}
\psfrag{f2/2}[cl][cl][\scalefig]{$\frac{1}{2}\hat{f}_{2}$}
\psfrag{f12XXXXXXXXXX}[cl][cl][\scalefig]{$f_2,\SNR_1=-1$ dB}
\psfrag{f1f2XXXXXXXXX}[cl][cl][\scalefig]{$\tilde{f}_{2},\SNR_1=-1$ dB}
\psfrag{f12XXXXXXXX}[cl][cl][\scalefig]{$f_2,\SNR_1=-0.6$ dB}
\psfrag{f1f2XXXXXXXX}[cl][cl][\scalefig]{$\tilde{f}_{2},\SNR_1=-0.6$ dB}
\psfrag{f3}[cl][cl][\scalefig]{$\hat{f}_{3}$}
\psfrag{f3/2}[cl][cl][\scalefig]{$\frac{1}{2}\hat{f}_{3}$}
\psfrag{f3/3}[cl][cl][\scalefig]{$\frac{1}{4}\hat{f}_{3}$}
\psfrag{f123XXXX}[cl][cl][\scalefig]{$f_3,\{1.5,3 \}$ }
\psfrag{f1f2f3XX}[cl][cl][\scalefig]{$\tilde{f}_{3},\{1.5,3 \}$ }
\psfrag{f123XX}[cl][cl][\scalefig]{$f_3,\{1.5,3.8 \}$ }
\psfrag{f1f2f3XXX}[cl][cl][\scalefig]{$\tilde{f}_{3},\{1.5,3.8 \}$}
\begin{center}
\scalebox{1}{\includegraphics[width=\linewidth]{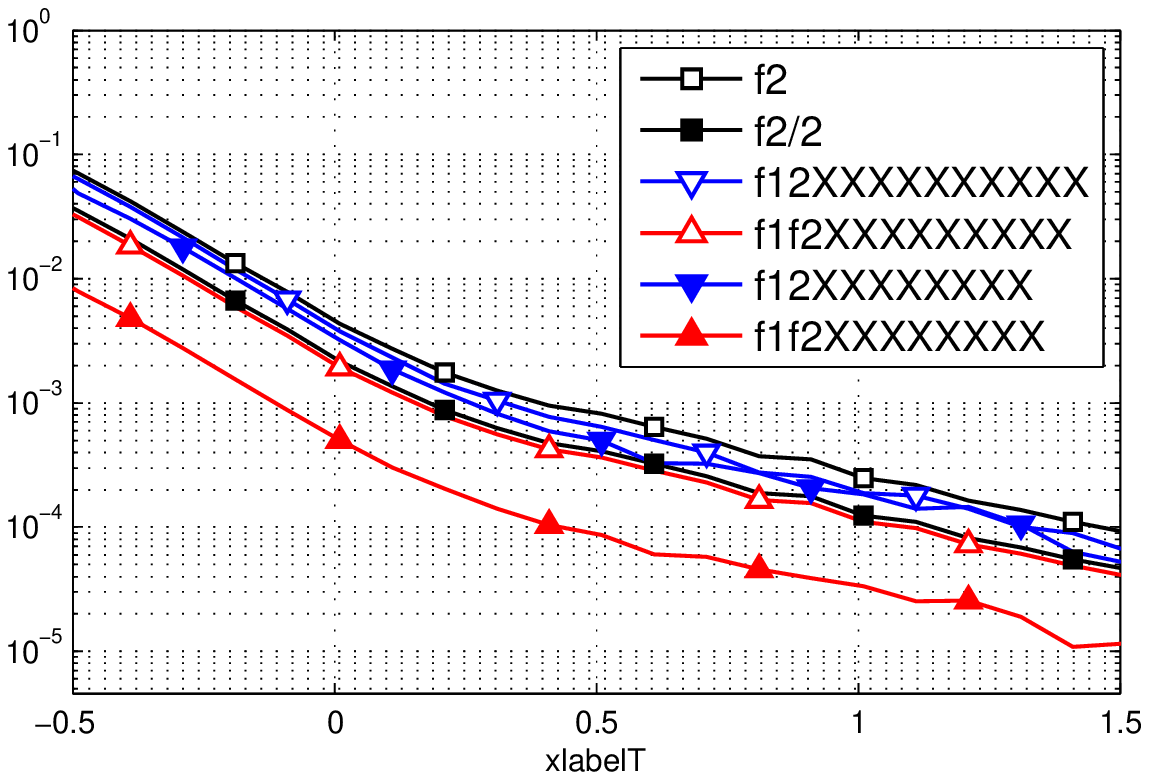}}\\
\scalebox{\scalefig}{a)}\\
\scalebox{1}{\includegraphics[width=\linewidth]{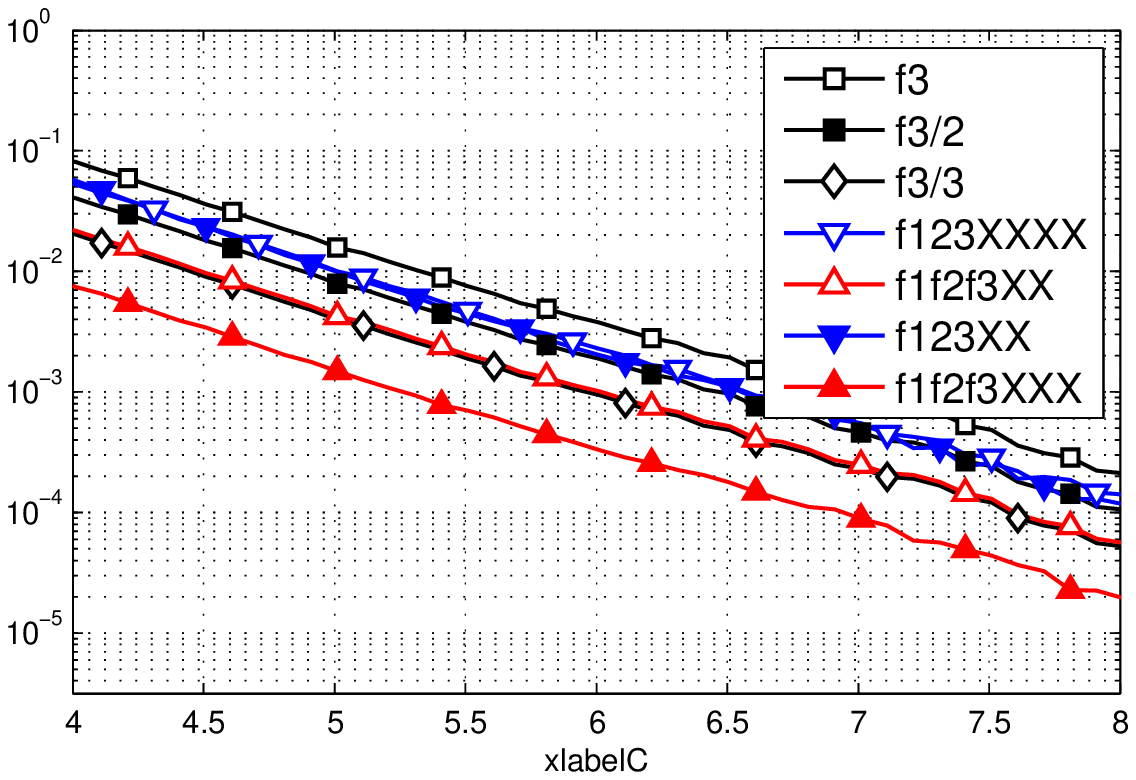}}\\
\scalebox{\scalefig}{b)}
\caption{Failure probabilities, $f_{\kk}$ \eqref{f.k}, and their approximations, $\hat{f}_{\kk}$ \eqref{E.cond.DE} and $\tilde{f}_{\kk}$ \eqref{E.cond.IE}, obtained  for  a)~$\kk=2$ with turbo code (4 decoding iterations), and b)~$\kk=3$ with convolutional encoder where the number in the braces denote \gls{snr}s in the first two rounds, \ie $\{\SNR_1,\SNR_2\}$~[dB].}\label{Fig:2trTurbo4It}
\end{center}
\end{figure}

\figref{Fig:2trTurbo4It} shows $f_k$, $\hat{f}_{\kk}$,  and $\tilde{f}_\kk$, for (a)~$\kk=2$, and (b)~$\kk=3$. For comparison we also show $\tfrac{1}{2}\hat{f}_{\kk}/2$, and $\frac{1}{2^{\kk-1}}\hat{f}_{\kk}$ (when $\kk=3$). We can observe that 
i)~$\tilde{f}_{\kk}$ can significantly underestimate $f_\kk$, particularly when $f_{l}, l<\kk,$ is small, which happens for relatively large values of $\SNR_{[l]}$, ii)~although $\hat{f}_{\kk}$ is an upper bound on $f_\kk$, both are very close and thus $\hat{f}_\kk$ should be preferred over $\tilde{f}_\kk$, and iii)~for $\kk=3$, the heuristic \gls{pep}-based bound $\check{f}_\kk=\tfrac{1}{2}\hat{f}_\kk$ provides a surprisingly accurate prediction of the results. This occurs most likely because the differences between the \gls{snr}s experienced in various rounds are not sufficiently large to attain the lower bound \eqref{property.3}.

\subsection{Application in i.i.d. block-fading}
Consider now the Rayleigh block-fading model where the \gls{snr}s in each transmission are modelled as \gls{iid} random variables $\SNRrv$ with distribution
\begin{align}\label{PDF_Rayleigh_inst_SNR}
\pdf_{\SNRrv}(x)=\frac{1}{\SNRav}\exp\biggl(-\frac{x}{\SNRav}\biggr),
\end{align}
where $\SNRav $ is the average SNR.

Our objective is to compare the IE and DE models in term of their applicability to predict the performance of the \gls{rrharq} which requires calculation of the average probability of the communication failure. We thus have to take expectation of  $f_\kk$ and $\tilde{f}_\kk$, with respect to the \gls{snr}s, $\SNRrv_1, \SNRrv_2, \ld, \SNRrv_\kk$, distributed according to \eqref{PDF_Rayleigh_inst_SNR}
\begin{align}
\label{f.avg.exact}
f^\tnr{avg}_\kk&=\Ex_{\SNRrv_{[1]}, \ld, \SNRrv_{[k]}}[f_k]\\
\label{f.avg.IE}
\tilde{f}^\tnr{avg}_\kk&=\Ex_{\SNRrv_{[1]}, \ld, \SNRrv_{[k]}}[\tilde{f}_k]
\end{align}
while, in the case of DE model we calculate
\begin{align}\label{f.avg.DE}
\hat{f}^\tnr{avg}_k&=\Ex_{\SNRrv_{[k]}}[\hat{f}_k]=
\int_{0}^{\infty} \PER(x) \pdf_{\SNRrv_{[k]}}(x) \dd x,
\end{align}
where $\SNRrv_{[k]}=\sum_{l=1}^k\SNRrv_{l}$, is a sum of \gls{iid} exponential variables and thus follows Gamma-distribution
\begin{align}\label{pdf.Naka}
\pdf_{\SNRrv_{[k]}}(x)= \frac{x^{k-1}}{(k-1)! \SNRav^k}  \exp(-\frac{x}{\SNRav}).
\end{align}

Thus, requiring only a one-dimensional integral \eqref{f.avg.DE}, the DE model provides  implementation advantage over the IE model, for which the explicit multi-dimensional integration \eqref{f.avg.IE} is needed.

This advantage can be then leveraged adopting a simple approximation of the decoding function used \eg in \cite{Liu04,Lagrange10}
\begin{align}\label{PER.SNR}
\PER(\SNR)
&=
\begin{cases}
1 &\text{if}\quad \SNR<\SNR_{\tr{th}}\\
\exp[-g(\SNR-\SNR_{\tr{th}}) ] &\text{if}\quad \SNR\geq\SNR_{\tr{th}}
\end{cases},
\end{align}
where the \emph{decoding threshold} $\SNR_{\tr{th}}$ and $g$ should be found from the empirical data using the curve fitting \cite{Liu04}.  

Integrating \eqref{PER.SNR} over the distribution \eqref{pdf.Naka} we obtain the following  closed form expression for $\hat{f}^\tnr{avg}_k$ in Rayleigh fading channel
\begin{align}
    \hat{f}^\tnr{avg}_k &= \frac{1}{(k-1)!} \big( \Gamma(k) -\Gamma(k, \frac{\SNR_{\tr{th}}}{\SNRav})+\nonumber \\
   &\qquad~\quad\exp(g\SNR_{\tr{th}})\frac{1}{(g\SNRav+1)^k} \Gamma(k, (g+\frac{1}{\SNRav})\SNR_{\tr{th}})\big ),
\end{align}
where $\Gamma(k,x)\triangleq\int_{x}^{\infty} \exp(-t) t^{k-1} \dd t $ and $\Gamma(k)\triangleq\Gamma(k,0)$ and are, respectively, the upper incomplete gamma function and, the gamma function.

Then, we can also calculate the average number of rounds, $\ov{K}$, of \gls{rrharq} with unlimited number of rounds, $K=\infty$
\begin{align}
\label{T.av}
\ov{K}&=\sum_{k=0}^{\infty} \hat{f}^\tnr{avg}_k = \int_{0}^{\infty} \PER(x) \sum_{k=0}^{\infty}  \pdf_{\SNRrv{[k]}}(x) \dd x \\
\label{T.av.2}
&= \int_{0}^{\infty} \mcL^{-1} \big[1/(1-P(s)),x\big] \PER(x) \dd x\\
\label{T.av.3}
&= \int_{0}^{\infty} \mcL^{-1} \big[1+1/(\SNRav s), x\big] \PER(x) \dd x\\ 
\label{T.av.4}
&= 1+(\SNR_{\tr{th}} + 1/g)/\SNRav,
\end{align}
where  $P(s)=\mcL[\pdf_\SNRrv(x), s]=1/(1+\SNRav s)$ is the Laplace transform of $\pdf_\SNRrv(x)$ and $\mcL^{-1}\big[P(s), x\big]$ is the inverse Laplace transform evaluated at $x$; to go from \eqref{T.av} to \eqref{T.av.2}, we used the geometric series $\sum_{k=0}^\infty P^k(s)=1/(1-P(s))$ as proposed in \cite{Larsson14}.

We note here that  \eqref{T.av.4} was also derived in \cite{Lagrange10}. Interestingly, however,  while \cite{Lagrange10} started from the IE model to calculate $\tilde{f}_k^\tr{avg}$, the numerous approximations applied to calculate the multi-dimensional integral \eqref{f.avg.IE} have lead to result we show in \eqref{T.av.4}. 

\begin{figure}[h]
\newcommand{\fsize}{0.9}
\psfrag{xlabel}[ct][ct][\fsize]{$\SNRav$[dB]}
\psfrag{ylabel}[ct][ct][\fsize]{$\PER(\SNR_{[1]}$}
\psfrag{f1k}[cl][cl][\fsize]{$\hat{f}^\tnr{avg}_{k}$}
\psfrag{f2k}[cl][cl][\fsize]{$\tilde{f}^\tnr{avg}_{k}$}
\psfrag{f1f2X}[cl][cl][\fsize]{$f^\tnr{avg}_{k}$}
\psfrag{k=2}[cl][cl][\fsize]{$k=2$}
\psfrag{k=3}[cr][cc][\fsize]{$k=3$}
\begin{center}
\scalebox{1}{\includegraphics[width=\linewidth]{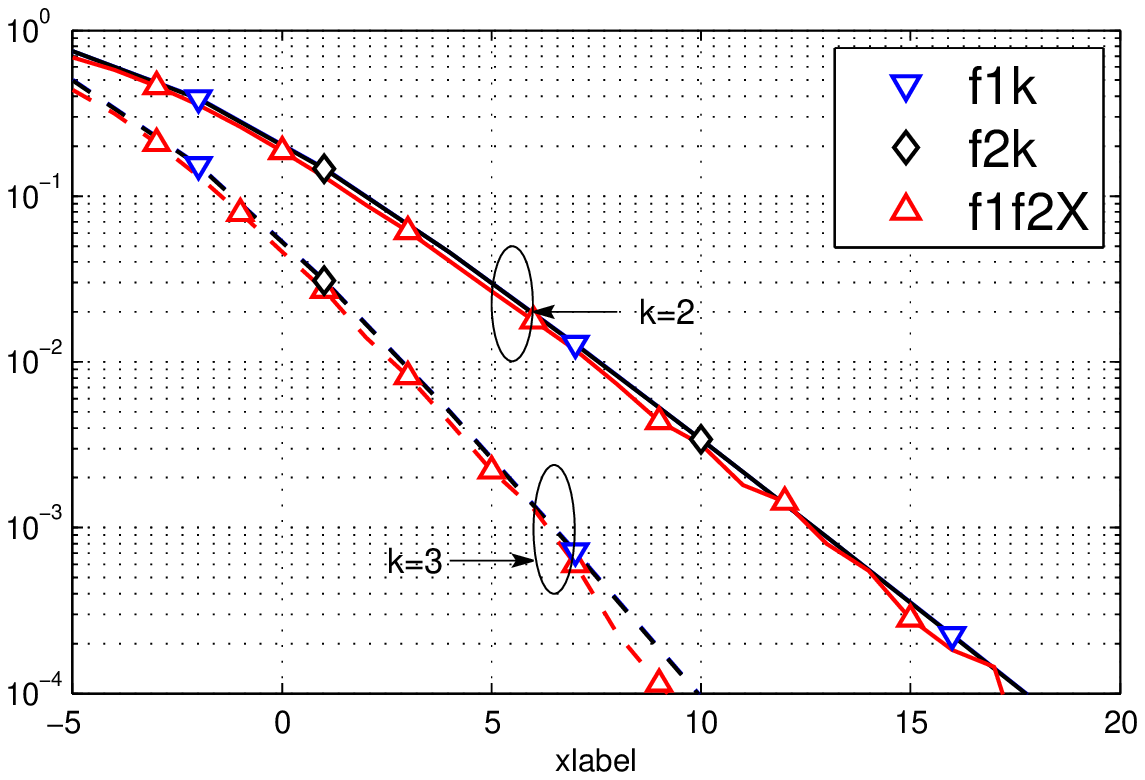}}
\caption{$\hat{f}^\tnr{avg}_{k}$, $\tilde{f}^\tnr{avg}_{k}$ and $f^\tnr{avg}_{k}$ for Raleigh fading channel as function of $\SNRav$ using turbo code (4 decoding iterations).}\label{Fig:Outage_Approx_fading_Turbo4It}
\end{center}
\end{figure}

\begin{figure}[h]
\newcommand{\fsize}{0.9}
\psfrag{xlabel}[ct][ct][\fsize]{$\SNRav$[dB]}
\psfrag{ylabel}[ct][ct][\fsize]{$\PER(\SNR_{[1]}$}
\psfrag{f1k}[cl][cl][\fsize]{$\hat{f}^\tnr{avg}_{k}$}
\psfrag{f2k}[cl][cl][\fsize]{$\tilde{f}^\tnr{avg}_{k}$}
\psfrag{f1f2X}[cl][cl][\fsize]{$f^\tnr{avg}_{k}$}
\psfrag{k=2}[cl][cl][\fsize]{$k=2$}
\psfrag{k=3}[cr][cc][\fsize]{$k=3$}
\begin{center}
\scalebox{1}{\includegraphics[width=\linewidth]{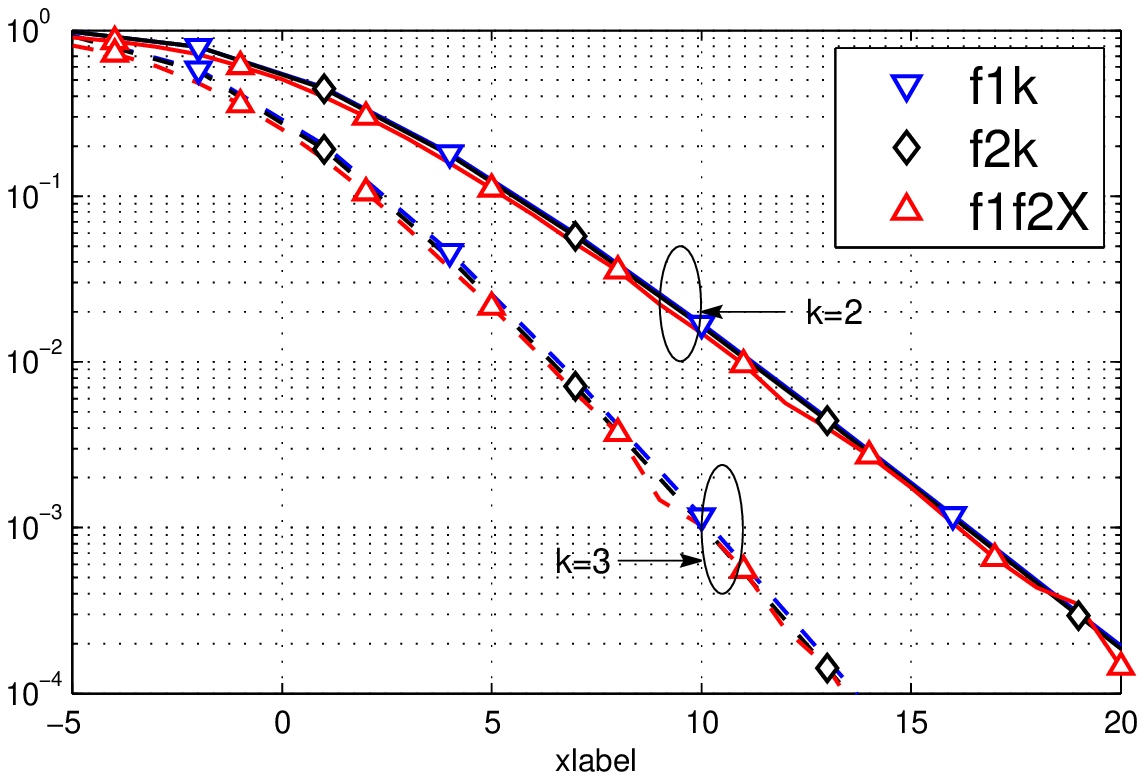}}
\caption{$\hat{f}^\tnr{avg}_{k}$, $\tilde{f}^\tnr{avg}_{k}$ and $f^\tnr{avg}_{k}$ for Raleigh fading channel as function of $\SNRav$ using convolutional code}\label{Fig:Outage_Approx_fading_Conv}
\end{center}
\end{figure}

\figref{Fig:Outage_Approx_fading_Turbo4It} and \figref{Fig:Outage_Approx_fading_Conv} show $f^\tnr{avg}_k$, $\hat{f}^\tnr{avg}_k$ and $\tilde{f}^\tnr{avg}_k$ ($k=2$ and $k=3$) where $\tilde{f}_k^\tr{avg}$ is obtained by numerical integration \eqref{f.avg.IE} and the actual value of $f^\tnr{avg}_k$ is obtained by simulations (implementation of \eqref{f.avg.exact} via Monte-Carlo integration). It turns out that, $f^\tnr{avg}_k$, $\hat{f}^\tnr{avg}_k$ and $\tilde{f}^\tnr{avg}_k$ are very similar, so the difference between IE and DE models is negligible in block-fading channels.  

This is not entirely surprising, because in fading channels the errors are mostly determined by the outage, \ie the event of $\set{\SNRrv<\SNR_\tnr{th}}$. Nonetheless, DE model is still attractive in this case because it provides exact closed form expression of the communication failure probability $\hat{f}_\kk$. 

\section{Conclusion}\label{Sec:Conclusions}
In this work, we proposed and analyzed a simple approximation to model the decoding errors in repetition redundancy (Chase combining) \gls{harq}. Our approximation uses solely the \gls{per} function of the receiver, is much more accurate than the expressions previously used in the literature, and it is a formal upper bound on the error-probability. Being equally simple as the alternative solutions and easier to manipulate, the expression we provided can be straightforwardly used to analyze \gls{harq} protocols, simulate the \gls{phy} behaviour, or adapt the transmission parameters.

\begin{appendices}
\section{Proof of \propref{Prop1}}\label{Appendix.proof}

To prove $P_{1:k}(d)\leq P_k(d)$, it is enough to apply the same bound as in  \eqref{Pr.nack.bound}.

We will now prove that $\frac{1}{2^{k-1}} P_k(d) \leq P_{1:k}(d)$.

Let  $\ba_l\T=[a_1,a_2,\ld,a_{l-1},a_l,0,\ld,0]\in \Real^\kk$ and $\bx\T=[x_1,x_2,\ld,x_\kk]\in \Real^\kk$ where
\begin{align}
a_l&=\sqrt{\SNR_{l}},\quad l=1,\ld,\kk
\end{align}
and
\begin{align}\label{}
x_l&=-\frac{\bd^\tr{H}\bz_l}{d},\quad l=1,\ld,\kk
\end{align}
is a zero-mean, unit-variance Gaussian random variable; here $\|\bd\|=d$. 

It is easy to see that $v_{[l]}=\ba_l\T\bx/\|\ba_l\|$, we can thus express the \gls{pep} \eqref{P.kk} as follows
\begin{align}
P_{l:\kk}(d)&=\Pr\{\ba_l\T\bx>\tfrac{1}{2}d\|\ba_l\|^2\wedge\ld \nonumber\\
            &\hskip 2cm \wedge\ba_k\T\bx>\tfrac{1}{2}d{\|\ba_k \|}^2\}.
\end{align}

We will use the intermediate Gaussian random variable $y_l=a_{l+1}\ba_l\T\bx-{\|\ba_l\|}^2x_{l+1}$, which has zero mean ($\Ex[y_l]=\Ex\big[ \ba_{l+1}\T\bx\big]=0$). 

\begin{lemma}\label{Lema.Ind}
The random variables $y_l$ and 
$\ba_{h}\T\bx$ are independent for $h=l+1,\ld,k$. 
\begin{proof}
for $h=l+1$ we have:
\begin{align}
\Ex[\ba_{l+1}\T\bx y_l]&=\Ex\big[(\ba_l\T\bx+a_{l+1}x_{l+1})\nonumber\\
&~~~~~~~~~~~~~~~~~~~(a_{l+1}\ba_l\T\bx-{\|\ba_l\|}^2x_{l+1})\big]\\
&=\Ex\big[a_{l+1}(\ba_l\T\bx)^2-a_{l+1}{\|\ba_l\|}^2x_{l+1}^2\big]\\
&=a_{l+1}{\|\ba_l\|}^2\Ex\big[x_1^2\big]-a_{l+1}{\|\ba_l\|}^2\Ex\big[x_{l+1}^2\big]\\
&=0.
\end{align}
Since $\ba_{l+1}\T\bx$ and $y_l$ are zero-mean Gaussian they are also independent.

For $h>l+1$, $\ba_{h}\T\bx=\ba_{l+1}\T\bx +\sum_{k=l+2}^h a_{k}x_k $. $\sum_{k=l+2}^h a_{k}x_k$ and $y_l$ are independent, therefore, it is obvious to see that $\ba_{h}\T\bx$ and $y_l$ are independent  $\forall h=l+2,\ld,k$. 
 \end{proof}
\end{lemma}
\begin{lemma}\label{evnt.reduction}
\begin{align}\label{}
&\set{\ba_{l+1}\T\bx>\tfrac{1}{2}d{\|\ba_{l+1}\|}^2 \wedge y_l>0}\nonumber\\
            &\hskip 3cm \implies \set{\ba_l\T\bx>\tfrac{1}{2}d{\|\ba_l\|}^2}.
\end{align}

\begin{proof}
We first multiply both sides of the inequality $\ba_{l+1}\T\bx>\tfrac{1}{2}d{\|\ba_{l+1}\|}^2$ with ${\|\ba_l\|}^2$, and the sides of the inequality $y_l>0$ with $a_{l+1}$. Then we  sum both and obtain
\begin{align}
{\|\ba_l\|}^2\ba_l\T\bx+a_{l+1}^2\ba_l\T\bx>\tfrac{1}{2}d{\|\ba_l\|}^2{\|\ba_{l+1}\|}^2,
\end{align}
which implies $\ba_l\T\bx>\tfrac{1}{2}d{\|\ba_l\|}^2$.
\end{proof}
\end{lemma}

Our objective is to show that
\begin{align}\label{Pk.bound}
P_{1:\kk}(d)\ge \frac{1}{2^{\kk-1}} P_{\kk}(d),
\end{align}
which hold due to the recursive application of the following.
\begin{lemma}\label{Lemm.main}
\begin{align}\label{Pk.bound2}
P_{l:\kk}(d) \ge \tfrac{1}{2} P_{l+1:\kk}(d), \quad l<\kk.
\end{align}
\begin{proof}
\begin{align}
P_{l:k}(d)&= \Pr\{\ba_l\T\bx>\tfrac{1}{2}d{\|\ba_l\|}^2 \wedge\ld  \nonumber\\
            &\hskip 2cm\wedge\ba_k\T\bx>\tfrac{1}{2}d{\|\ba_k\|}^2\}\label{eq1}\\
           &\ge\Pr\{y_l>0\wedge\ba_{l+1}\T\bx>\tfrac{1}{2}d{\|\ba_{l+1}\|}^2 \wedge\ld \nonumber\\
           &\hskip 2cm \wedge\ba_k\T\bx>\tfrac{1}{2}d{\|\ba_k\|}^2\}\label{eq2}\\
           &=\tfrac{1}{2}\Pr\{\ba_{l+1}\T\bx>{\|\ba_{l+1}\|}^2 \wedge\ld \nonumber\\
           &\hskip 2cm 
           \wedge\ba_k\T\bx>\tfrac{1}{2}d{\|\ba_k\|}^2\}\label{eq3}\end{align}
where to go from \eqref{eq1} to \eqref{eq2} we used \lemref{evnt.reduction} and to go from \eqref{eq2} to \eqref{eq3} we used \lemref{Lema.Ind}, and the fact that $y_l$ is a zero-mean, Gaussian random variable so $\PR{y_l>0}=\tfrac{1}{2}$. 
\end{proof}
\end{lemma}

\end{appendices}

\end{document}